\documentclass[twocolumn,twocolappendix]{aastex631}

\usepackage{amssymb}
\usepackage{amsmath}
\usepackage[]{graphicx}
\usepackage{enumerate}
\usepackage{subeqnarray}
\usepackage{cases}
\usepackage{mathrsfs,amssymb}

\shorttitle{Bombardment of CO ice by cosmic rays: ...}
\shortauthors{Ivlev et al.}

\tightenlines

\begin{document}

\title{Bombardment of CO ice by cosmic rays:\\ I. Experimental insights into the microphysics of molecule destruction and sputtering}

\author{Alexei V. Ivlev} \affiliation{Max-Planck-Institut f\"ur Extraterrestrische Physik, D-85748 Garching, Germany }
\author{Barbara M. Giuliano} \affiliation{Max-Planck-Institut f\"ur Extraterrestrische Physik, D-85748 Garching, Germany }
\author{Zoltán Juhász} \affiliation{Institute for Nuclear Research (Atomki), PO Box 51, H-4026 Debrecen, Hungary}
\author{Péter Herczku} \affiliation{Institute for Nuclear Research (Atomki), PO Box 51, H-4026 Debrecen, Hungary}
\author{Béla Sulik} \affiliation{Institute for Nuclear Research (Atomki), PO Box 51, H-4026 Debrecen, Hungary}
\author{Duncan V. Mifsud} \affiliation{Centre for Astrophysics and Planetary Science, School of Physics and Astronomy, University of Kent, Canterbury CT2 7NH, United Kingdom} \affiliation{Institute for Nuclear Research (Atomki), PO Box 51, H-4026 Debrecen, Hungary}
\author{Sándor T.S. Kovács} \affiliation{Institute for Nuclear Research (Atomki), PO Box 51, H-4026 Debrecen, Hungary}
\author{K.K. Rahul} \affiliation{Institute for Nuclear Research (Atomki), PO Box 51, H-4026 Debrecen, Hungary}
\author{Richárd Rácz} \affiliation{Institute for Nuclear Research (Atomki), PO Box 51, H-4026 Debrecen, Hungary}
\author{Sándor Biri} \affiliation{Institute for Nuclear Research (Atomki), PO Box 51, H-4026 Debrecen, Hungary}
\author{István Rajta} \affiliation{Institute for Nuclear Research (Atomki), PO Box 51, H-4026 Debrecen, Hungary}
\author{István Vajda} \affiliation{Institute for Nuclear Research (Atomki), PO Box 51, H-4026 Debrecen, Hungary}
\author{Nigel J. Mason} \affiliation{Centre for Astrophysics and Planetary Science, School of Physics and Astronomy, University of Kent, Canterbury CT2 7NH, United Kingdom} \affiliation{Institute for Nuclear Research (Atomki), PO Box 51, H-4026 Debrecen, Hungary}
\author{Sergio Ioppolo} \affiliation{School of Electronic Engineering and Computer Science, Queen Mary University of London, London E1 4NS, United Kingdom} \affiliation{Centre for Interstellar Catalysis (InterCat), Department of Physics and Astronomy, Aarhus University, DK 8000 Aarhus, Denmark}
\author{Paola Caselli} \affiliation{Max-Planck-Institut f\"ur Extraterrestrische Physik, D-85748 Garching, Germany }


\correspondingauthor{A.~V.~Ivlev} \email{ivlev@mpe.mpg.de}

\begin{abstract}
We present a dedicated experimental study of microscopic mechanisms controlling radiolysis and sputtering of astrophysical ices due to their bombardment by cosmic ray ions. Such ions are slowed down due to inelastic collisions with bound electrons, resulting in ionization and excitation of ice molecules. In experiments on CO ice irradiation, we show that the relative contribution of these two mechanisms of energy loss to molecule destruction and sputtering can be probed by selecting ion energies near the peak of the electronic stopping power. We have observed a significant asymmetry, both in the destruction cross section and the sputtering yield, for pairs of ion energies corresponding to same values of the stopping power on either side of the peak. This implies that the stopping power does not solely control these processes, as usually assumed in the literature. Our results suggest that electronic excitations represent a significantly more efficient channel for radiolysis and, possibly, also for sputtering of CO ice. We also show that the charge state of incident ions as well as the rate for CO$^+$ production in the ice have  negligible effect on these processes.
\end{abstract}



\section{Introduction}

Subrelativistic cosmic rays (CRs) play a crucial role in the evolution of dense astrophysical environments. Such CRs are known to be the primary source of gas ionization and heating in dense cores of molecular clouds, thus controlling dynamical and chemical processes that accompany practically all stages of star formation \citep{McKee1989, Keto2008, Dalgarno2006, Glassgold2012}. Apart from the dynamical and structural evolution of dense cores and the formation of disks, these processes also include the growth of icy mantles on the surface of dust grains, ice processing, and chemical reactions. The gas-phase chemistry is inherently linked to the chemistry occurring within the icy mantles \citep{Bergin1995, Caselli1999}, as the bombardment by CRs affects the structural and chemical properties of ices and leads to desorption of chemical species back to the gas phase \citep{Leger1985, Hasegawa1993, Vasyunin2017}.

While the gas-phase chemistry and the physical processes induced by CRs in gas are reasonably well understood, the phenomena occurring in icy mantles are far more complex. The energy deposited by CRs upon their collisions with dust grains causes structural evolution, compaction, and desorption of ices \citep{Brown1982, Johnson1982, Leto2003, Loeffler2005, Palumbo2006, Rothard2017, Dartois2013, Dartois2021}. Also, CRs induce nonequilibrium chemistry in ices, and recently developed astrochemical models include reactions between excited species as well as high-temperature reactions \citep{Shingledecker2019, Shingledecker2020, Kalvans2019, Garrod2019, Anders2019, Gronoff2020, ODonoghue2022}. Nevertheless, one should keep in mind that the energy of CRs is primarily transferred to the bound electrons, causing excitation of the electronic states of molecules and their ionization, and there are diverse possibilities of how this deposited energy may then drive the physical and chemical processes in the ice.

Ionization and electronic excitation by CRs are the two well-understood channels of energy deposition \citep{Ziegler1985, Ziegler1988, Ziegler2010}. However, their individual contributions into various physical and chemical processes driven in ices are still poorly studied. The principal aim of the present paper is to experimentally explore the relative importance of electronic excitation and ionization for processing of astrophysical ices by CRs. Specifically, we have conducted dedicated experiments on a CO ice to investigate the relative efficiency of the two mechanisms of energy deposition for destruction of CO molecules and their sputtering. In Section~\ref{motivation} we discuss how the contribution of these two mechanisms to radiolysis and sputtering can be probed by selecting ion energies near the peak of the electronic stopping power. In Section~\ref{exp} we describe the experimental procedure and techniques that provide complementary data on the energy dependencies of the CO destruction cross section and sputtering yield. Section~\ref{results} contains results of the measurements, showing that both radiolysis and sputtering are significantly enhanced at lower ion energies, where the proportion of energy deposited through electronic excitations is expected to be higher. Implications of the results for microscopic processes that may potentially be involved in radiolysis and sputtering as well as the astrophysical implications of our work are discussed in Section~\ref{discussion}, and a summary is presented in Section~\ref{conclusions}.

\section{Motivation}
\label{motivation}

For projectile ions with $E_{\rm ion}\gtrsim 10$~keV/u, their energy loss due to elastic collisions with nuclei of target molecules (``nuclear'' stopping power) becomes less important than that due to inelastic collisions with the bound electrons \citep{Ziegler1985, Ziegler1988, Ziegler2010}. The resulting energy loss rate, associated with the ionization of molecules and excitation of their electronic states, is commonly called electronic stopping power, $S_{\rm el}(E_{\rm ion})$.

While ice processing by low-energy ions occurs due to the elastic collisions \citep{Fama2008}, the electronic stopping power is the key parameter which controls characteristics of radiation chemistry and sputtering of ices at energies relevant to CRs \citep[see, e.g.,][and references therein]{Rothard2017}. For a given ice composition, $S_{\rm el}$ is considered to solely determine the sputtering yield and radiolysis cross sections \citep{Brown1980, Brown1982, Johnson1982, Loeffler2005, SeperueloDuarte2009, SeperueloDuarte2010, Pilling2010, Mejia2015b, deBarros2020, Dartois2018, Dartois2021}. There are a number of ways in which the energy transferred to electrons due to ionization and excitation can be converted into the kinetic energy of molecules, and how the molecules can be destroyed. The ionized molecules can undergo barrierless reactions with nearby molecules \citep{Herbst1973, Watson1974}, or they can be destroyed in dissociative recombination with electrons \citep{Dalgarno1976}, converting the repulsive energy of dissociation products into their kinetic energy. Also, mutual electrostatic repulsion between ions that are created along the projectile track may break the intermolecular bonds, and accelerate ions and molecules in a ``Coulomb explosion'' \citep{Brown1984}. At the same time, excitation of dissociative electronic states \citep{Cosby1993, Jamieson2006} -- directly by ions or by the ejected electrons -- also leads to molecular destruction and acceleration. Furthermore, excited molecules can efficiently react with their neighbors: e.g., reaction between excited and ground-state CO molecules produces CO$_2$ and atomic carbon \citep{Jamieson2006}.

The importance of these microscopic processes for radiolysis and sputtering of ices is largely unknown. One cannot exclude that different processes may play a dominant role in different ices and for different ion energies. For radiolysis, there have been attempts to parameterize the efficiency of these processes in terms of mean energies lost by ions per ionization and excitation \citep{Shingledecker2018a, Shingledecker2018b}. In the following section we discuss how the relative efficiency of ionization and excitation channels can be probed in experiments.

\subsection{Ionization versus excitation}
\label{versus}

\begin{figure}
\begin{center}
\includegraphics[width=.95\columnwidth]{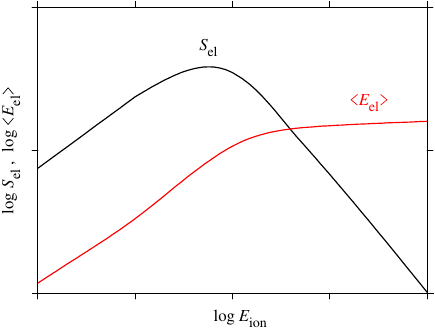}
\caption{The electronic stopping power of ions, $S_{\rm el}$, and the mean kinetic energy of ejected electrons, $\langle E_{\rm el}\rangle$, versus the ion kinetic energy $E_{\rm ion}$. The plot illustrates the generic behavior near the stopping power peak, the axes ticks indicate decades. Exact values depend on the projectile ion and target molecule (see text for details), the peak is typically reached at $E_{\rm ion}$ between $\sim0.1$~MeV/u and $\sim1$~MeV/u. The plotted curves represent protons in molecular hydrogen: $S_{\rm el}$ is obtained from the National Institute of Standards and Technology (NIST) database \url{http://physics.nist.gov/Star}, $\langle E_{\rm el}\rangle$ is computed for the energy distribution of \citet{Rudd1992}.}
\label{fig_S_and_E}
\end{center}
\end{figure}

For high ion energies, corresponding to the decreasing branch of the stopping power curve in Figure~\ref{fig_S_and_E}, $S_{\rm el}(E_{\rm ion})$ is described by the classical Bethe formula \citep{Bethe1932}. For non-relativistic ions, it has the universal dependence $S_{\rm el}(E_{\rm ion})\propto \ln\mathcal{V}/E_{\rm ion}$, where $\mathcal{V}\gg1$ is the dimensionless ion velocity,
\begin{equation*}
\mathcal{V}=\sqrt{\frac{m_{\rm el}}{m_{\rm ion}}\frac{E_{\rm ion}}I}\;,
\end{equation*}
with the normalization by the mean excitation energy $I$ of electronic shells of the target molecule. The maximum of $S_{\rm el}(E_{\rm ion})$ is reached where the ion velocity becomes comparable to the characteristic velocity of bound electrons in the molecule \citep{Rudd1992}. For different ions and molecules, this roughly corresponds to the value of $\mathcal{V}$ of the order of unity \citep{Ziegler2010}; this value slowly increases with the ion atomic number $Z$, changing from $\sim1$ for protons (maximum at $E_{\rm ion}\sim0.1$~MeV) to $\sim3$ for iron ions (maximum at $E_{\rm ion}\sim1$~MeV/u). We note that the electronic stopping power completely dominates over the nuclear contribution (by a factor of $\sim$~300--1000, see \url{http://physics.nist.gov/Star}) at energies near and above the peak of $S_{\rm el}(E_{\rm ion})$.

\begin{figure}
\begin{center}
\includegraphics[width=\columnwidth]{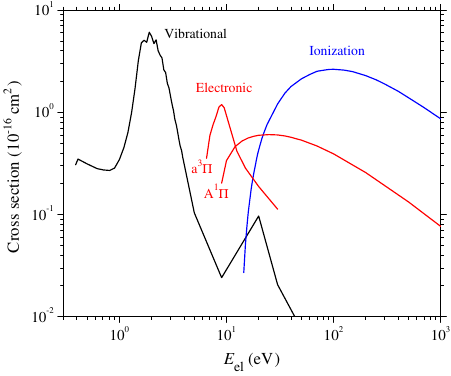}
\caption{Cross sections for inelastic electron collisions with CO molecules. For electronic excitations, the two main contributions from the a$^3\Pi$ and A$^1\Pi$ states are shown. The data are taken from \citet{Itikawa2015}.} \label{fig_x_sections}
\end{center}
\end{figure}

The electrons generated in ionizing collisions have a certain energy spectrum and, hence, they are able to produce secondary ionization and excitation. The energy probability distribution of these ejected electrons, $p(E_{\rm el})$, critically depends on the magnitude of $\mathcal{V}$ \citep{Rudd1988, Rudd1992}. In the Bethe regime $\mathcal{V}\gg1$, we have $p(E_{\rm el})\propto E_{\rm el}^{-2}$ and the maximum electron energy is limited by a large value of $\approx4\mathcal{V}^2I$. The shape of the distribution weakly changes with $E_{\rm ion}$ in this case, and the mean energy of ejected electrons, $\langle E_{\rm el}\rangle=\int E_{\rm el}\:p(E_{\rm el})\:dE_{\rm el}$, approaches a constant value of the order of a few $I$. On the other hand, for $\mathcal{V}\ll1$ the ejected electrons are localized at lower energies, and their distribution has a sharp exponential dependence,
\begin{equation*}
p(E_{\rm el})\sim \frac{\alpha}{\mathcal{V}I}\exp\left(-\frac{\alpha E_{\rm el}}{\mathcal{V}I}\right),
\end{equation*}
where $\alpha$ is an order-of-unity constant (dependent on the target molecule). Thus, the mean energy of ejected electrons in this regime is $\langle E_{\rm el}\rangle\sim \mathcal{V}I$, i.e., it approximately increases as $\propto\sqrt{E_{\rm ion}}$. A typical dependence $\langle E_{\rm el}\rangle$ versus $E_{\rm ion}$ is plotted in Figure~\ref{fig_S_and_E}.

Figure~\ref{fig_x_sections} shows the main cross sections for electron-impact inelastic collisions with CO molecules \citep{Itikawa2015}. For energetic ions with $E_{\rm ion}$ above 0.3--1~MeV/u, corresponding to the Bethe regime $\mathcal{V}\gg1$, the mean energy of ejected electrons exceeds 20--30~eV and therefore supports secondary ionization. By reducing $E_{\rm ion}$ below the stopping power peak, we can lower $\langle E_{\rm el}\rangle$ and, thus, increase the role of other inelastic collisions. In particular, we can make electronic excitations the dominant mechanism of energy loss by the electrons. It is noteworthy that, in this case, direct electronic excitations by ions will play an increasingly important role in their energy loss, too: experimental measurements of the so-called ``$W$-value'' of ions in various gases (that is, the mean energy lost by an ion per ionization) commonly show that $W$ becomes substantially higher as $E_{\rm ion}$ decreases below the stopping power peak \citep{Srdoc1995}.

The above analysis suggests that, by changing $E_{\rm ion}$ in the vicinity of the stopping power peak, the energy deposited by ions in ice can be distributed in varying proportions between ionization and electronic excitations. Therefore, by measuring CO destruction and sputtering on the two sides of the peak -- yet for same values of the stopping power -- we should be able to probe the relative efficiency of ionization and excitation for these processes.

\section{Experimental}
\label{exp}

All experiments discussed in this paper have been conducted at the Institute for Nuclear Research (Atomki) located in Debrecen, Hungary.

\subsection{Ice preparation and ion beams}
\label{prep}

\begin{table*}
\label{wtf}
  \caption{{\bf Experimental parameters}$^{\rm a}$ {\bf and results}$^{\rm b}$}
  \begin{center}
  \begin{tabular}{ c | c | c | c | c | c || c }
        \hline
        \rule[-1.5ex]{0pt}{4.5ex}
  Ion & $E_{\rm ion}$ & $S_{\rm el}$ & $N_0$ & $\Delta F$ & $F_{\rm max}$ & $\sigma_{\rm des}+Y_{\rm sp}/N_0$ \\
      & (MeV) & ($10^{-15}$~eV~cm$^2$/CO) & ($10^{17}$~cm$^{-2}$) & ($10^{12}$~ion/cm$^2$) & ($10^{14}$~ion/cm$^2$) & ($10^{-15}$~cm$^2$) \\
        \hline\hline
        \rule[-.5ex]{0pt}{3.ex}
  He$^+$    & 0.3  & 84.6           & 7.18 & \it{0.587} & 0.549 & 1.57\\
            &      &                & 7.32 & \it{0.493} & 0.553 & 1.66\\
            &      &                & 6.32$^{\rm d,e}$ & \it{0.471} & 0.507 & 1.38\\
            &      &                & 6.87 & 5.03  & 1.00 & 1.69\\
            &      &                & 6.83 & 5.00  & 1.00 & 1.73\\
        \cline{2-7}
        \rule[-.5ex]{0pt}{3.ex}
            & 0.5  & 94.8           & 5.87$^{\rm e}$ & 5.11  & 0.508 & 1.63\\
            &      &                & 7.18$^{\rm d}$ & 5.23  & 1.00  & 1.84\\
        \cline{2-7}
        \rule[-.5ex]{0pt}{3.ex}
            & 0.75 & 94.8           & 6.31$^{\rm e}$ & 5.25  & 0.564 & 1.42\\
            &      &                & 7.67 & 4.97  & 1.00 & 1.45\\
        \cline{2-7}
        \rule[-.5ex]{0pt}{3.ex}
            & 1.0  & 90.3           & 6.42$^{\rm e}$ & 4.92  & 0.554 & 1.19\\
            &      &                & 7.60 & 4.87  & 1.00 & 1.25\\
        \cline{2-7}
        \rule[-.5ex]{0pt}{3.ex}
            & 1.5  & 77.2           & 5.79$^{\rm e}$ & \it{0.651}  & 0.507 & 0.932\\
            &      &                & 6.14$^{\rm e}$ & \it{0.491}  & 0.504 & 1.04\\
            &      &                & 6.89 & 5.08  & 1.00 & 0.961\\
            &      &                & 6.90 & 4.97  & 1.00 & 0.813\\
        \cline{2-7}
        \rule[-.5ex]{0pt}{3.ex}
            & 2.0  & 67.2           & 5.85 & 5.38  & 1.00 & 0.699\\
            &      &                & 6.45 & 4.98  & 1.01 & 0.634\\
        \hline
        \rule[-.5ex]{0pt}{3.ex}
  He$^{2+}$ & 2.0  & 67.2           & 5.88 & 4.87  & 1.00 & 0.673\\
            &      &                & 6.52 & 6.26  & 1.01 & 0.677\\
        \cline{2-7}
        \rule[-.5ex]{0pt}{3.ex}
            & 6.0  & 32.5$^{\rm c}$ & 7.32 & 5.01  & 1.00  & 0.229\\
            &      &                & 7.49 & 5.05  & 0.999 & 0.229\\
        \hline
        \rule[-.5ex]{0pt}{3.ex}
  H$^+$     & 0.2  & 28.1$^{\rm c}$ & 7.38 & 5.16  & 1.00 & --\\
            &      &                & 7.73 & 5.27  & 1.00 & --\\
        \hline
  \end{tabular}
  \end{center}
$^{\rm a}$The stopping power $S_{\rm el}(E_{\rm ion})$ of He ions in CO gas, measured by \citet{Bourland1971} for ion energies $E_{\rm ion}$ between 0.3~MeV and 2~MeV; the initial column density of CO ice film $N_0$, measured in each experiment before irradiation; the ion fluence step $\Delta F$, and the maximum accumulated fluence $F_{\rm max}$ in each experiment. The fluence steps were kept approximately constant in regular experiments; the {\it fine-step} experiments, {\it marked in italics}, started with shorter steps (the indicated values), and continued with longer steps as fluence accumulated. Experimental uncertainties are summarized in Section~\ref{error}.

$^{\rm b}$The destruction cross section $\sigma_{\rm des}$, computed from FTIR measurements. Shown are the mean values of $\sigma_{\rm des}+Y_{\rm sp}/N_0$ for the experiments with He ions, plotted in Figure~\ref{fig_slope} ($Y_{\rm sp}$ is the sputtering yield); mean values for the two  experiments with protons cannot be reliably computed (see Section~\ref{CO_column}). Destruction dominates over sputtering at each energy (see Section~\ref{destruction}).

$^{\rm c}$Estimated values of $S_{\rm el}$ for CO: derived from the NIST database (\url{http://physics.nist.gov/Star}) for CO$_2$ and O$_2$ by using Bragg's rule $S_{\rm CO}=S_{\rm CO_2}-\frac12S_{\rm O_2}$.

$^{\rm d}$Two experiments with large systematic errors of $\approx10\%$ in the ion fluence (see Section~\ref{error}).

$^{\rm e}$Six experiments where the QMS sensitivity was not measured directly (see Section~\ref{error}).

\vspace{.2cm}
\end{table*}

The Ice Chamber for Astrophysics-Astrochemistry (ICA) has been employed for the growth and processing of the CO ice samples. The ICA is an ultra-high-vacuum chamber which hosts a closed-cycle cryostat with up to four IR-transparent substrates mounted on a copper holder, and thus allows for the production of ice analog replicas created under identical experimental conditions. The position of the sample holder can be controlled using a vertical linear manipulator, and its tilt angle with respect to the beam is set by a 360$^{\circ}$ rotation stage. Once the sample holder is cooled down to a minimum temperature of 20~K, ice samples can be deposited onto the substrates via background deposition of dosed gases. Further details on the ICA experimental facility can be found in \citet{Herczku2021} and \citet{Mifsud2021}. The ice processing is then conducted using a 2~MV Tandetron accelerator \citep{Biri2021}, capable of delivering a variety of high-energy ions into the ICA, to simulate CR bombardment.

In Table~1 we summarize key parameters of all experiments whose analysis is presented below in Section~\ref{results}. For our studies, pure CO ice films are deposited at 20~K on top of three ZnSe substrates, which allows us to perform three experiments per deposition process. The deposition lasts 8~min at an average rate of just below 0.05~$\mu$m/min (as estimated from evolving IR spectra, see Section~\ref{FTIR}). The pressure of the chamber is $\approx8\times 10^{-6}$~mbar during deposition and $\approx5\times 10^{-9}$~mbar after deposition. A final check is carried out before starting experiments, to ensure that the beam has the desired current and homogeneity.

All experiments consist of a series of irradiation steps (see Table~1; a regular fluence step lasts about 14~s). Each step is immediately followed by recording of a IR spectrum, and the sequence repeats itself until the accumulated fluence of $\sim10^{14}$~ion/cm$^2$ is reached (which takes approximately 1~hr in each experiment). In some cases, the same beam energy is used for two consecutive experiments, and then the second experiment starts immediately after the end of the first one. To change the beam energy, additional time between 30~min and 1.5~hr is required (depending on the new desired energy).

Irradiation in all experiments is performed using beam currents between 100~nA and 150~nA over an area of 1.13~cm$^2$, with projectile ions impacting the ice films at angles of 36$^\circ$ to the normal. Under such conditions, we estimate the temperature increase during irradiation to be always much smaller than $0.1$~K. Further details can be found in \citet{Herczku2021}.

Homogeneity of the ion beam with an indirectly measured fluence $F$ at the irradiated area is provided by the beam guiding system, described in detail in \citet{Herczku2021} (see their Figure~2). The current is continuously measured and integrated in a monitoring Faraday cup (F1), with a beam-size-defining collimator at its bottom. One movable Faraday cup (F2) is used to directly measure the current transmitted by F1 to the sample, before and after each irradiation step. In addition, a third Faraday cup (F3) is used to check the beam homogeneity at the beginning and in the end of each experiment.

\subsection{FTIR spectroscopy}
\label{FTIR}

Spectroscopic changes occurring in the ice films upon their irradiation are monitored using a Thermo Nicolet Nexus 670 FTIR spectrophotometer, with a spectral range of 4000--650~cm$^{-1}$ (2.5--15.4~$\mu$m) and a nominal resolution of 1~cm$^{-1}$. The spectra are collected using 128 scans at a rate of 1 scan per second. The IR beam, orthogonal to the ZnSe substrates, is detected in transmission by a mercury-cadmium-telluride detector. The entire IR beam path and the detector are kept under a purified air flow, to prevent absorption by water and carbon dioxide in the air.

\begin{figure}
\begin{center}
\includegraphics[width=\columnwidth]{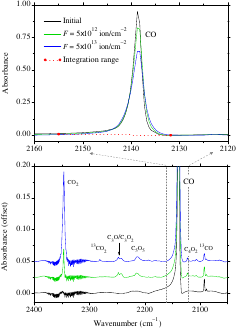}
\caption{An example of FTIR spectra, measured in an experiment with 0.3~MeV He$^+$ ions before irradiation and at two different ion fluences. The upper panel shows zoom-in on the CO absorption band at 2139~cm$^{-1}$, the band area is computed by integrating it between 2155--2132~cm$^{-1}$ (indicated by the red bullets). The lower panel illustrates the evolution in a broader range of 2050--2400~cm$^{-1}$, showing several radiolysis products forming in the ice.}
\label{fig_FTIR}
\end{center}
\end{figure}

Figure~\ref{fig_FTIR} illustrates the evolution of IR spectra with the ion fluence in our experiments. The CO column density $N$ is calculated in a standard way \citep[e.g.,][]{SeperueloDuarte2009, Pilling2010}, from the band area of a vibrational absorption feature. The column density is $N=\int \tau_\nu d\nu/A$, where $\tau_\nu$ is optical depth and $A$ is the absorption band strength. The absorbance measured by the FTIR spectrometer is $\tau_\nu/\ln10$, and therefore the band area is $S=\int \tau_\nu d\nu/\ln10$. This allows us to determine the ice column density for a given fluence by computing $S(F)$, provided the band strength versus fluence is known,
\begin{equation}\label{N(F)}
N(F)=\ln10\;\frac{S(F)}{A(F)}\;.
\end{equation}
We use the most intense absorption feature of CO ice at 2139~cm$^{-1}$, assuming $A_0=8.7\times 10^{-18}$~cm/CO molecule for the initial band strength before irradiation \citep{GonzalesDiaz2022}. The way to disentangle and determine dependencies $A(F)$ and $N(F)$ is discussed in Section~\ref{CO_column}.

The initial column densities of CO ice films for each experiment are listed in Table~1. We assume $\rho_{\rm CO}\approx 0.87$~g/cm$^3$ for the initial mass density of CO ice \citep{Luna2022, GonzalesDiaz2022}, which corresponds to the ice number density $n_{\rm CO}\approx1.8 \times 10^{22}$~cm$^{-3}$ and a monolayer thickness $n_{\rm CO}^{-1/3} \approx3.8$~\AA. The initial thickness of ice films, estimated as $N_0/n_{\rm CO}$, varies between $\approx0.3$ and $\approx0.4$~$\mu$m for all experiments.

The measured IR spectra allow us to determine concentrations of some residual molecules deposited on the ice surface between the end of CO deposition and the beginning of irradiation. Apart from CO$_2$ pollution, whose spectral feature near 2346~cm$^{-1}$ is evident in the initial spectrum (see the lower panel of Figure~\ref{fig_FTIR}), also OH stretching bond of water molecules (near 3250~cm$^{-1}$, not shown) is detected in many experiments. Using the bands strengths of $7.6\times 10^{-17}$~cm/molecule for CO$_2$ \citep{Jamieson2006} and $2.2\times 10^{-16}$~cm/molecule for the amorphous compact water \citep{Leto2003}, the maximum estimated column densities are about $2\times10^{15}$~cm$^{-2}$ for CO$_2$ and $\lesssim10^{16}$~cm$^{-2}$ for water. Thus, the resulting pollution is substantially less than 1\% for CO$_2$ and about 1\% for water in the worst case.

\subsection{QMS measurements}
\label{QMS}

The quadrupole mass spectrometer (QMS) of the type Pfeiffer QME200 is continuously running in each experiment, to analyze residual gases in the chamber. Before and during the ice deposition, the mass range from 1 to 50 amu is monitored with 1 amu steps, to check the composition of deposited gases. Before starting irradiation, the mass range is reduced to 27--29 amu in order to monitor the sputtered CO molecules of mass 28. The shortest possible time steps are used (so that transient effects at the beginning and in the end of the mass scan can still be avoided), which allows us to sample the CO concentration every 1.3~s and, thus, resolve the sputtering dynamics during the irradiation steps.

\begin{figure}
\begin{center}
\includegraphics[width=\columnwidth]{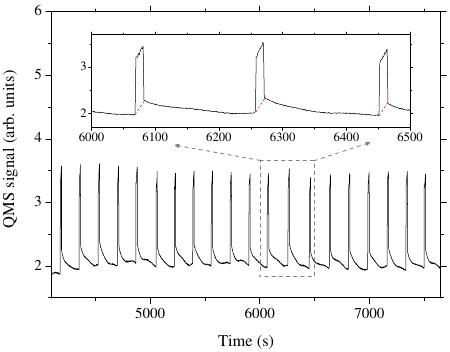}
\caption{An example of the QMS signal, recorded for 0.3~MeV He$^+$ ions in an experiment with regular fluence steps of $\Delta F\sim5\times10^{12}$~ion/cm$^2$ (see Table~1). Zoom-in on the three consecutive irradiation steps illustrates the shape of the resulting signal pulses. The sputtering yield is estimated at each step as the pulse area (bounded by the red dashed line) divided by the fluence step.}
\label{fig_QMS}
\end{center}
\end{figure}

In Figure~\ref{fig_QMS}, the current measured for mass 28  by a Faraday cup of the QMS is shown for one of experiments with regular fluence steps. The beginning and the end of each irradiation step are clearly visible, producing a sharp signal pulse relative to the background level. Long tails follow these pulses, while during the pulse the current monotonically increases. (The maximum increase of the CO partial pressure, measured with the QMS during irradiation by He ions, is between $5\times10^{-9}$ and $7\times10^{-9}$~mbar.) This behaviour is likely due to a finite response time of the pumping system, unable to quickly reach equilibrium after a sudden change in the load of CO molecules. The integral of each pulse is computed relative to the background line, which spans the first and the last points of the pulse (the red dashed lines in Figure~\ref{fig_QMS}). To estimate the sputtering yield for each irradiation step, the corresponding integral is divided by the fluence step. The resulting values measured during each experiment are then used to compute the average and the statistical error for a given ion energy. Since the response to a given load of CO molecules has not been calibrated, the measurements do not give us the absolute values of the sputtering yield, but its relative dependence on the energy.

We note that the sputtering yield for the first irradiation steps often exhibits rather strong fluctuations. In some experiments, the sputtering yield shows a marginal trend of a slight (within 10--20\%) increase at $F\lesssim 10^{13}$~ion/cm$^2$; for larger fluences, it remains constant on average, showing no statistically significant dependence on $F$. We are uncertain whether this weak increase in the beginning may be attributed to ice compaction (see Section~\ref{CO_column}), or to other possible minor effects (e.g., sputtering of polluting molecules deposited on the ice surface between the end of CO deposition and the beginning of irradiation, see Section~\ref{FTIR}). Furthermore, for {\it fine-step} experiments, the starting pulses are too short to provide good statistics, while the monotonic increase of the signal during longer irradiation steps (at larger accumulated fluences) leads to its saturation, which noticeably affects the resulting values of the pulse area. To exclude these undesirable effects and, simultaneously, avoid any bias in the analysis of {\it fine-step} and regular experiments, we use a common fluence range between $1\times 10^{13}$~ion/cm$^2$ and $5\times 10^{13}$~ion/cm$^2$ to compute the sputtering yield.

\subsection{Experimental uncertainties}
\label{error}

Uncertainties in the ion fluence affect both FTIR and QMS results. The statistical error in measuring the currents in the three Faraday cups is negligible compared to beam fluctuations and beam misalignment. Assuming them to be the main sources of error in the fluence, systematic uncertainties are estimated from the measured values of the currents. Except for two experiments with large systematic errors of 10\% (indicated in Table~1: one is due to beam fluctuations, the other is due to misalignment), the estimated systematic uncertainties are below 5\% and, in most cases close to 2\%. Only one experiment is excluded from the analysis (a {\it fine step} experiment at 1.5~MeV, not shown in Table~1) because of unstable beam conditions at a level of 30\%. As expected, no systematic dependence on the beam current value is observed neither in FTIR nor in QMS results.

The QMS measurements in different experiments (repeated for same ion energy) show variations which are substantially larger than the statistical error of each individual experiment. Apart from the fluence uncertainties mentioned above, the QMS measurements may have other sources of systematic errors, such as ice impurities and variations in the QMS sensitivity. The most significant contribution to ice impurities stems from N$_2$ molecules, having the same mass as CO. The QMS measurements of partial pressures in the chamber before deposition show that the residual abundance of N$_2$ is usually a factor of 2--3 higher than the water abundance. Since the resulting water pollution of ice deduced from IR measurements does not exceed 1\% (see Section~\ref{FTIR}), we can safely estimate N$_2$ pollution to be within a few percent. The QMS sensitivity is checked every day, by comparing the pressure readout at the ionization gauge and the measured QMS signal during the ice deposition (the repeatability of the pressure measurements is 5\%). The derived sensitivity is always found to be smaller than the reference value in the beginning of the experimental series (by several percent), changing irregularly from day to day. These daily values are then used to compensate for the reduction of sensitivity. In six experiments (also indicated in Table~1), where the QMS signal could not be recorded during the ice deposition, the expected sensitivity reduction is compensated by assuming the average reduction value from the other experiments (which is about 6\%).

\section{Results}
\label{results}

In this section we present and analyze results of FTIR and QMS measurements that were performed for the experiments listed in Table~1. This analysis allows us to determine the destruction cross section of CO molecules as well as the CO sputtering yield. We show that both parameters are significantly enhanced at lower ion energies, where electronic excitations are expected to have a higher relative contribution to the stopping power.

\subsection{CO column density versus ion fluence}
\label{CO_column}

Consider a general case of irradiation of pure ice composed of molecules M. Radiolysis products forming at early irradiation stages are not yet abundant enough to substantially affect the M sputtering, while destruction of the products cannot yet contribute appreciably to the backward formation of M. In this case, we can expect the following simple equation \citep[see, e.g.,][]{SeperueloDuarte2009, Pilling2010} to accurately describe how the column density $N$ of molecules M evolves with the ion fluence $F$:
\begin{equation}\label{dN/dF}
\frac{dN}{dF}=-\sigma_{\rm des}N-Y_{\rm sp}\;,
\end{equation}
where $\sigma_{\rm des}$ is the cross section of molecular destruction, occurring in pure bulk ice per ion impact, and $Y_{\rm sp}$ is the sputtering yield of the molecules, i.e., the number of molecules M ejected per ion impact from the surface of pure ice.


The analytical solution of Equation~(\ref{dN/dF}) is straightforward. As we are aiming to analyze early stages of irradiation, where variations in $N(F)$ are still small, it is sufficient to use a linear expansion of the solution over $F$. Substituting the result for $N(F)/N_0$ in Equation~(\ref{N(F)}), we obtain
\begin{equation}\label{S_norm}
\frac{S(F)}{S_0}\approx\frac{A(F)}{A_0}\left[1-\left(\sigma_{\rm des}+\frac{Y_{\rm sp}}{N_0}\right)F\right],
\end{equation}
where $N_0$, $S_0$, and $A_0$ denote the values measured before irradiation. In order to relate the measured band area $S(F)$ to the parameters $\sigma_{\rm des}$ and $Y_{\rm sp}$, we need to exclude the unknown dependence $A(F)$ describing the band strength evolution. This can be done in a general way by keeping in mind that variations in $A$ are caused by the ice compaction -- a (relatively) very fast process, evolving at fluences/doses which are typically much smaller than those needed for a substantial reduction of the column density \citep{Leto2003, Loeffler2005, Palumbo2006, Urso2016, Dartois2013, Dartois2021}. We can therefore assume that the band strength reaches some asymptotic value $A_\infty$, characteristic of compact ice, well before approaching the fluence range relevant to our
analysis of $N(F)$.

\begin{figure}
\begin{center}
\includegraphics[width=\columnwidth]{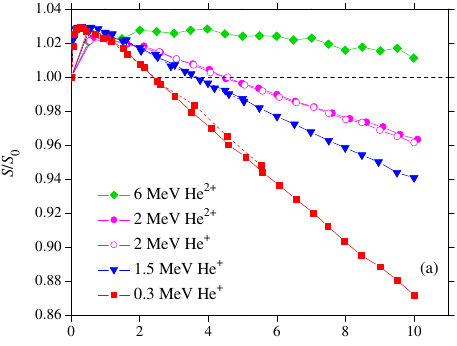}
\includegraphics[width=\columnwidth]{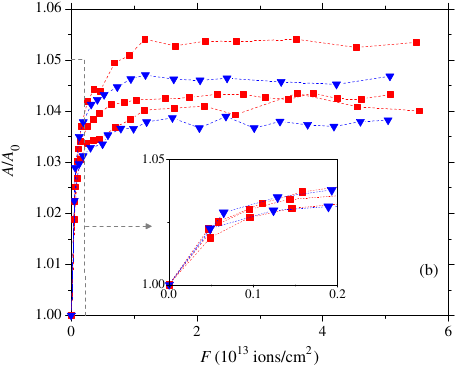}
\caption{(a)~The area of CO absorption band normalized to its initial value, $S/S_0$, versus the fluence $F$ of He ions. Results for several different energies and charges of incident ions are illustrated. The symbols connected by the solid lines represent experiments with regular fluence steps of $\Delta F\sim5\times 10^{12}$~ion/cm$^2$, the dashed lines are for the {\it fine-step} experiments with shorter steps of $\Delta F\sim5\times 10^{11}$~ion/cm$^2$ at the beginning (see Table~1). (b)~The normalized band strength, $A/A_0$ versus $F$, for the {\it fine-step} experiments at 0.3~MeV and 1.5~MeV. The inset shows the evolution at the starting fluence steps.} \label{fig_band}
\end{center}
\end{figure}

Figure~\ref{fig_band}(a) illustrates several measurements of the CO band area versus the fluence of He ions, performed for
different ion energies. Uncertainties of the measurements (not shown in the plot) are determined by the systematic errors in
the fluence (see Section~\ref{error}). The measured points connected by the solid lines represent our regular experiments,
carried out with approximately constant fluence steps, while the dashed lines show the {\it fine-step} experiments,
performed for 0.3~MeV and 1.5~MeV He ions with short starting steps (see Table~1). All measurements show a sudden increase
of the normalized area $S(F)/S_0$ at $F<10^{13}$~ion/cm$^2$, with a maximum excess over unity of about 3\%, while at higher
fluences a transition to a linearly decreasing $S(F)/S_0$ is observed. This rapid increase in the band area reflects the
corresponding increase in the band strength $A$, occurring due to ice compaction \citep{Mejia2015, Mejia2022}, while the
observed linear decrease at higher fluences justifies the use of Equation~(\ref{S_norm}).

We note that the comparative measurements performed with 0.2~MeV protons exhibit a similar behavior. The stopping power of the protons is only 15\% smaller than that of 6~MeV He$^{2+}$ ions, and the measured dependencies $S(F)/S_0$ (not shown) decrease somewhat slower than the green curve in Figure~\ref{fig_band}(a). However, due to a slower compaction in case of 0.2~MeV protons, the common fluence range chosen for all experiment is too short to reliably disentangle $A(F)$ and $N(F)$ in the experiments with protons.

By fitting the results obtained for He ions at higher fluences with a linear function, $S(F)/S_0=a-bF$, and comparing it with Equation~(\ref{S_norm}), we have 
$\sigma_{\rm des}+Y_{\rm sp}/N_0=b/a$. Hence, the data illustrated in Figure~\ref{fig_band}(a) allow us to derive a combined measure of CO destruction and sputtering for a given ion energy. In particular, we notice a significant difference for 0.3~MeV and 1.5~MeV ions, even though the stopping power values for the two energies are different by only $\approx10\%$. In Section~\ref{destruction} we present detailed analysis of these results. Furthermore, we point out that the measurements for singly and doubly charged 2~MeV ions are practically indistinguishable. In Section~\ref{charge} we discuss expected effects of the charge state of incident ions on CO radiolysis and sputtering.

The band strength evolution, $A(F)/A_0$, can be reconstructed from the {\it fine-step} experiments with 0.3~MeV and 1.5~MeV ions. It is obtained by dividing the measured $S(F)/S_0$ by the factor in the brackets in Equation~(\ref{S_norm}), i.e., by $1-(b/a)F$. The results are presented in Figure~\ref{fig_band}(b). As expected \citep{Mejia2015, Mejia2022}, the band strength rapidly evolves to the asymptotic value $A_\infty$ which is some 4\% higher than the initial value $A_0$. In contrast to the significantly different behavior seen for the two energies at higher fluences, in Figure~\ref{fig_band}(a), the dependencies $A(F)/A_0$ are very similar: they all ultimately saturate at $F\gtrsim10^{13}$~ion/cm$^2$, which corresponds to the deposited dose $S_{\rm el}F$ of around 0.3--1~eV/molecule. On the other hand, from the inset it is evident that the main change in $A(F)$ occurs at even lower fluences, well below the minimum fluence step. Therefore, we estimate the characteristic dose, needed for a substantial compaction of CO ice, to be as low as a few times $10^{-2}$~eV/molecule.

\subsection{CO destruction}
\label{destruction}

\begin{figure}
\begin{center}
\includegraphics[width=\columnwidth]{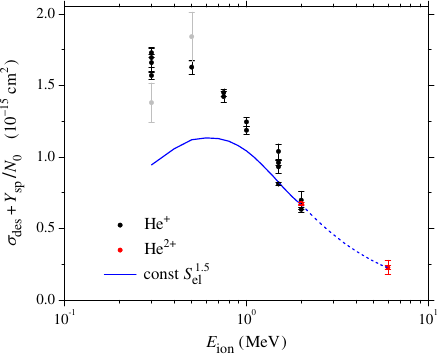}
\caption{The value of $\sigma_{\rm des}+Y_{\rm sp}/N_0$ derived for different ion energies from the analysis of CO absorption band (the mean values for each experiment are listed in Table~1). The destruction cross section $\sigma_{\rm des}(E_{\rm ion})$ is estimated to dominate for all probed energies (see Section~\ref{destruction}). The blue line represents the scaling dependence $\sigma_{\rm des}\propto S_{\rm el}^{1.5}$ suggested in the literature (see text for details); the part of it shown by the dashed line is estimated from the NIST database (see Table~1). The data for 2~MeV as well as for 6~MeV He$^{2+}$ ions (2 red symbols each) practically overlap. The light-gray symbols indicate the two experiments with large systematic errors in the ion fluence (see Section~\ref{error} and Table~1).} \label{fig_slope}
\end{center}
\end{figure}

Figure~\ref{fig_slope} summarizes the calculated mean values of $\sigma_{\rm des}+Y_{\rm sp}/N_0$ for all experiments with He ions (also listed in Table~1). The error bars show the combined standard deviation for the statistical and systematic errors. The statistical error was calculated as the unbiased sample variance for the linear fit of $S(F)/S_0$ (see Section~\ref{CO_column}), assessment of the systematic error for the fluence $F$ is described in Section~\ref{error}. Leaving aside the two experiments with large systematic errors, we see that the resulting values monotonically increase as $E_{\rm ion}$ decreases. Let us first find out what contribution -- molecule destruction or sputtering -- dominate the results.

Experiments by \citet{Dartois2021} and \citet{SeperueloDuarte2010} on irradiation of CO ice by heavy ions suggest that, for their conditions \citep[see Table~1 in][]{Dartois2021}, the destruction and sputtering terms were of the same order of magnitude. Assuming the stopping power scaling $Y_{\rm sp}\propto S_{\rm el}^2$ for the sputtering yield \citep[see][and references therein]{Rothard2017} and $\sigma_{\rm des}\propto S_{\rm el}^{1.5}$ for the destruction cross section \citep{deBarros2020}, we can extrapolate the ratio of the two terms to the conditions of our experiments, which yields $\sigma_{\rm des}N_0/Y_{\rm sp}\sim3$ for the maximum stopping power in our Table~1. We conclude that the results plotted in Figure~\ref{fig_slope} primarily represent the destruction cross section of CO molecules.

Now we can probe applicability of the scaling relation $\sigma_{\rm des}\propto S_{\rm el}^{1.5}$ suggested in the literature. The blue line in Figure~\ref{fig_slope} shows this dependence, confirming that it indeed provides good description of the data for ion energies well above the stopping power peak. However, for $E_{\rm ion}\lesssim1.5$~MeV this dependence systematically underestimates the destruction cross section. The discrepancy increases steadily as the peak is approached, and becomes about 80\% for the minimum probed energy.

Hence, for similar values of $S_{\rm el}(E_{\rm ion})$ on either side of the stopping power peak, the efficiency of destruction is significantly higher for a lower ion energy -- where the proportion of energy deposited through electronic excitations is higher, as discussed in Section~\ref{versus}. This suggests that electronic excitations in CO ice are a significantly more efficient radiolysis channel than ionization. Such a conclusion is further corroborated by the fact that a similar significant asymmetry is also observed for almost all radiolysis products detected in our experiments (including those indicated in Figure~\ref{fig_FTIR}). These results will be presented and discussed in a separate paper, focused on the radiolysis chemistry.

\subsection{CO sputtering}
\label{sputtering}

\begin{figure}
\begin{center}
\includegraphics[width=\columnwidth]{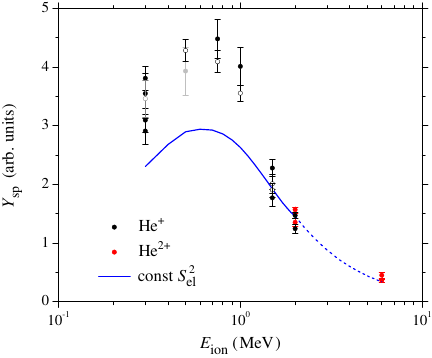}
\caption{Energy dependence of the CO sputtering yield $Y_{\rm sp}$, obtained from uncalibrated QMS measurements. The blue line shows the scaling dependence $Y_{\rm sp}\propto S_{\rm el}^2$ suggested in the literature (see text for details); the part of it shown by the dashed line is estimated from the NIST database (see Table~1). The light-gray symbols indicate the two experiments with large systematic errors in the ion fluence, and the open symbols show the six experiments where the QMS sensitivity was not measured directly, but the average sensitivity from the other experiments is used (see Section~\ref{error} and Table~1).}
\label{fig_yield}
\end{center}
\end{figure}

QMS detection of CO molecules emitted from the ice surface during irradiation provides an independent and complementary source of information, allowing us to measure the energy dependence of the sputtering yield. Figure~\ref{fig_yield} shows results of uncalibrated QMS measurements that were performed for all experiments listed in Table~1 for He ions. As in Figure~\ref{fig_slope}, the error bars show the combined standard deviation for the statistical and systematic errors. Despite a stronger data scatter than in the FTIR measurements, the results demonstrate a similar overall trend as $E_{\rm ion}$ changes across the stopping power peak. At energies well above the peak, the commonly used scaling dependence $Y_{\rm sp}\propto S_{\rm el}^2$ goes through the data points, while for $E_{\rm ion}\lesssim1.5$~MeV it substantially underestimates the measurements. The sputtering yield exhibits a jump by about 50\%, and at lower energies its variation approximately follows the amplified blue line. Similar to the enhancement of the destruction cross section, the substantial enhancement of $Y_{\rm sp}$ suggests that electronic excitations may be a more efficient channel for CO sputtering, too. Of course, a possible interplay of different microscopic processes may also contribute to the increased sputtering. Additional experiments and theoretical work are certainly needed to fully understand these results.

It is noteworthy that the data points depicted in Figures~\ref{fig_slope} and \ref{fig_yield} suggest different shapes for the respective energy dependencies. This fact can be used as an additional argument to reasonably exclude a possible substantial contribution of the sputtering term $Y_{\rm sp}/N_0$ in the dependence plotted in Figure~\ref{fig_slope}. Indeed, the values of $N_0$ (averaged for each energy) vary by less than 10\% between different energies, as it follows from Table~1. Therefore, if the sputtering term were playing an important role in Figure~\ref{fig_slope}, we would see a trend to reach a maximum near 0.75~MeV in that figure, instead of the observed monotonic behavior.

Figure~\ref{fig_yield} shows that CO sputtering, similar to CO destruction, is rather insensitive to the charge of incident ions: the difference between average mean values of $Y_{\rm sp}$ for doubly and singly charged 2~MeV ions is about 8\%. This fact and its possible implications are discussed in Section~\ref{charge}.

Finally, we mention that the sputtering measurements performed for 0.2~MeV protons (not shown in Figure~\ref{fig_yield}) are completely consistent with the measurements for He ions at higher energies. The sputtering yield obtained for 0.2~MeV protons (the average mean value) is $\approx20\%$ smaller than that for 6~MeV He$^{2+}$ ions. This is an expected result, since the peak of the proton stopping power
is located at $E_{\rm ion}\approx95$~keV \citep{Ziegler2010}, and thus the dependence $Y_{\rm sp}\propto S_{\rm el}^2$ (which predicts a $\approx30\%$ smaller $Y_{\rm sp}$ for protons than for the He ions) should be applicable for protons too.

\section{Discussion}
\label{discussion}

The results presented in Section~\ref{results} show that both the destruction cross section and sputtering yield of CO ice for a given value of the stopping power are significantly larger at lower ion energies, below the stopping power peak. As explained in Section~\ref{versus}, electronic excitations of CO molecules play a higher role in the energy deposition at lower ion energies, and therefore we conclude that excitations are a significantly more efficient channel for radiolysis and, possibly, also for sputtering than ionization. One can speculate that excitations of dissociative electronic states, simultaneously leading to destruction of molecules and acceleration of dissociation products, may be the microscopic process behind the observed results.

In our case, we can reasonably rule out other microscopic processes that have been previously discussed to contribute to radiolysis and sputtering. One of such processes could be production of ionized molecules along the track of a projectile ion \citep{Johnson1982, Brown1984}. Their dissociative recombination may simultaneously lead to the destruction and acceleration of the products, similar to the dissociative excitation. Data on cross sections for ionization and charge changing in CO are only available for a limited range of He ion energies \citep{Rudd1985b, Rudd1985a}, but the results practically coincide with the cross sections measured in N$_2$. Therefore, we estimate the effective cross section for CO$^+$ production by He ions from the results computed for N$_2$ \citep[plotted by the solid line in Figure~11 of][]{Bug2013}, showing that the effective cross section remains constant within measurement errors for $0.3$~MeV~$\lesssim E_{\rm ion}\lesssim1$~MeV. If CO$^+$ production were the main process behind the observed destruction and sputtering, then both $\sigma_{\rm des}(E_{\rm ion})$ and $Y_{\rm sp}(E_{\rm ion})$ would be approximately constant in that energy range, too -- however, the qualitative behavior seen in Figures~\ref{fig_slope} and \ref{fig_yield} is obviously different. For the same reason, we can also exclude sputtering due to ``Coulomb explosion'' \citep{Bringa2003, Iza2006, Pilling2010}, a process associated with the Coulomb repulsion between neighboring CO$^+$ ions produced along the track.

Another process that had been invoked in earlier publications \citep{Brown1982, Johnson1982} to explain enhancement of $Y_{\rm sp}(E_{\rm ion})$ at energies below the stopping power peak are charge-changing collisions. Electron capture or loss by incident ions affect the energy deposition in upper ice layers, before the ions acquire equilibrium distribution of the charge states \citep{Fama2007, Assmann2017}. If this process were playing an important role in our experiments, we would see a significant difference in the sputtering yields by 2~MeV He$^{2+}$ and He$^+$ ions, but this is clearly not the case in Figure~\ref{fig_yield}. In the following Section~\ref{charge} we elaborate on effects associated with the charge of incident ions.

\subsection{Charge of incident ions}
\label{charge}

Comparison between the experiments performed with He$^+$ and He$^{2+}$ ion beams at 2~MeV indicates practically no effect of the charge of incident ions on CO destruction, whereas the difference between the sputtering yields is estimated to not exceed 10\%. To understand these results, we recall that the reference values of $S_{\rm el}(E_{\rm ion})$ are obtained by measuring an {\it average} rate of energy loss by ions moving through a given target material. This yields an {\it equilibrium} stopping power, corresponding to an equilibrium distribution of charge states of the projectile ions in the material. This implies that the energy deposition in ice near the surface may significantly deviate from $S_{\rm el}(E_{\rm ion})$ if the charge of incident ions noticeably deviates from the mean equilibrium charge \citep{Fama2007, Assmann2017}.

For the energy range probed in our experiments, He$^+$ are the dominant equilibrium ions up to about 0.6~MeV; for higher energies, these are He$^{2+}$, with the fraction above 90\% at $E_{\rm ion}=2$~MeV \citep{Bug2013}. Hence, in the experiments with 2~MeV He$^{2+}$ ions the energy deposition is practically constant and equal to $S_{\rm el}(E_{\rm ion})$ along the entire ion track, but for He$^+$ ions it is substantially reduced in upper ice layers, because the stopping power of He$^+$ at such energies is a factor of two lower than that of He$^{2+}$ \citep{Johnson1982}. The depth of the charge equilibration is determined by the cross sections of charge changing between the two states, $\sigma_{12}(E_{\rm ion})$ and $\sigma_{21}(E_{\rm ion})$. We were not able to find the cross sections measured at 2~MeV in CO, but the comparison of cross sections available for various gases at lower energies shows that the measured values for CO coincide with those for N$_2$ within experimental errors, both for charge changing and ionization \citep{Rudd1985b, Rudd1985a}. Using the values measured in N$_2$ at 2~MeV, we obtain $\sigma_{12}\approx9\times10^{-17}$~cm$^2$ and $\sigma_{21}\approx5\times10^{-18}$~cm$^2$ \citep{Bug2013}.

In Appendix~\ref{A1} we show that the charge equilibration depth is set by the value of $\sigma_\Sigma= \sigma_{12}+ \sigma_{21}$, which is about $10^{-16}$~cm$^2$ in our case. It is then not surprising that the measurements shown in Figures~\ref{fig_band}(a) and \ref{fig_slope} for singly and doubly charged ions are indistinguishable: indeed, molecules are destroyed along the entire ion track in the ice film, whose initial column density $N_0$ is a factor of $\sim50$ larger than the value of $\sigma_\Sigma^{-1}$. However, the situation is more subtle for the sputtering yield, presented in Figure~\ref{fig_yield}. The sputtering occurs from upper ice layers, limited along the track by the so-called sputtering depth $N_{\rm sp}$. Recent measurements of $N_{\rm sp}$ by \citet{Dartois2021}, performed in CO and CO$_2$ for different values of the stopping power (see their Figures~7 and 8), suggest an approximately linear dependence $N_{\rm sp}\propto S_{\rm el}$. For 2~MeV He ions in CO, their measurements give $N_{\rm sp}\sim 10^{16}$~cm$^2$, i.e., $\sigma_\Sigma N_{\rm sp}\sim1$ and therefore our sputtering results require further analysis. Let us focus on it.

The average measured value of $Y_{\rm sp}$ for He$^{2+}$ is only $\sim10\%$ higher than that for He$^+$. From this we conclude that the {\it effective} (viz., determining the sputtering) values of the stopping power for the two ions should be very close. The dependence $Y_{\rm sp}\propto S_{\rm el}^2$, which still appears to work at $E_{\rm ion}=2$~MeV, suggests that the two effective values may deviate by some 5\%. Generally, the effective stopping power must depend on the product $\sigma_\Sigma N_{\rm sp}$: If the sputtering depth is much smaller than the charge equilibration depth ($\sigma_\Sigma N_{\rm sp}\ll1$), we expect the effective value to be the stopping power of the incident-charge ions (this would be $\approx0.5S_{\rm el}$ for He$^+$ and $\approx S_{\rm el}$ for He$^{2+}$), while in the opposite limit $\sigma_\Sigma N_{\rm sp}\gg1$ the effective value must tend to $S_{\rm el}$. Obviously, in our case we are dealing with a situation which is closer to the latter limit.

The effective value of the stopping power should be mostly determined by the total energy deposited in ice within the sputtering depth. We can then reasonably assess the effective value by averaging the energy deposition along the ion track from zero to $N_{\rm sp}$. However, assuming $\sigma_\Sigma N_{\rm sp}=1$ in Equation~(\ref{N_Sbar}) of Appendix~\ref{A1} yields the average stopping power for He$^+$ ions which is still $\sim30\%$ smaller than $S_{\rm el}$. As the deviation from equilibrium decreases as $\propto (\sigma_\Sigma N)^{-1}$, this implies that $N_{\rm sp}$ in our case may be a factor of $\sim10$ larger than the value suggested by \citet{Dartois2021}.

Based on the above analysis, we can generalize the results of our experiments to other ices and conclude that, as long as $\sigma_\Sigma N_0\gg1$, the measured $\sigma_{\rm des}(E_{\rm ion})$ is not expected to depend significantly on the charge of incident ions. On the other hand, $Y_{\rm sp}(E_{\rm ion})$ is expected to reveal a significant dependence for less volatile ices, such as H$_2$O and CO$_2$, where the sputtering depth (for a given value of $S_{\rm el}$) is much smaller than for CO \citep{Dartois2018, Dartois2021}, and thus the product $\sigma_\Sigma N_{\rm sp}$ can be very small.

\subsection{Astrophysical implications}

Dust grains in cold ($\sim10$~K) dense cores of molecular clouds are covered with thick icy mantles composed of hundreds of molecular monolayers \citep[e.g.,][]{Draine1985, Oeberg2011, Kalvans2015, ChaconTanarro2019, Caselli2022}. The existing gas-grain chemistry models suggest that the mantle structure is largely controlled by a combination of the solid-phase processes of water formation and destruction as well as by the adsorption and desorption of CO molecules \citep{Watanabe2004, Garrod2008, Cuppen2017}. The inner layers of icy mantles form at earlier evolutionary stages and primarily contain water, but catastrophic freeze-out of CO on dust grains \citep{Caselli1999, Goldsmith2001, Jorgensen2005}, occurring when the gas number density reaches over $\sim10^4$~cm$^{-3}$ while the temperature is below 10~K, leads to the formation of upper layers predominantly composed of CO ice \citep{Oeberg2011, Kalvans2015, Vasyunin2017}. CO-rich ices are also expected in the midplane of protoplanetary disks, at the location of the CO snow line \citep[e.g.,][]{Qi2013}.

There are multiple microscopic processes driven by CRs in astrophysical ices that can lead to their radiolysis and sputtering. All such processes are initiated either by ionization or electronic excitation of ice molecules, and the results of our experiments allow us to understand the relative importance of these two channels of energy deposition in the upper, CO-dominated layers of icy mantles. We should, however, keep in mind that different microscopic processes may play a dominant role in different ices, and therefore further dedicated experiments with other main components of icy mantles need to be conducted in the future (in particular, with less volatile components, such as H$_2$O and CO$_2$, to probe the effect of the binding energy).

The conclusion that electronic excitations are a more efficient channel for radiolysis and, possibly, also for sputtering makes it reasonable to suggest that excitations of dissociative states may be the microscopic process behind the observed results. Thus, our experiments can provide important constraints for available physical models of ices undergoing CR bombardment, by specifying the mechanism of energy deposition. Furthermore, the reported results will also significantly improve the overall predictability of existing astrochemical models. \citet{Shingledecker2018a} and \citet{Shingledecker2018b} have recently developed a model of CR-driven solid-state chemistry, where radiolysis processes in ices are assumed to be completely associated with electronically excited ``suprathermal'' species. The reaction rates in their model are expressed via the efficiency of ionization and excitation which, in turn, are parameterized in terms of estimated mean energies lost per single event by projectile ions  \citep[see, e.g.,  Equations~4 through 12 in][]{Shingledecker2018b}. Combining our results presented in Figure~\ref{fig_slope} with detailed analysis of $W$-values for He ions \citep{Srdoc1995} and inelastic cross sections for electrons in CO \citep{Itikawa2015} can deliver explicit energy dependencies for the reaction efficiencies and, thus, provide the precise recipe for the ``suprathermal'' chemical network implemented in \citet{Shingledecker2018a} and \citet{Shingledecker2018b}.

Microscopic mechanisms of interaction of CRs with astrophysical ices are qualitatively distinct from those operating in case of UV or x-ray photons. CRs colliding with bound electrons of ice molecules lose their energy in relatively small portions, which leads to efficient ionization and electronic excitation along the CR tracks. The proportion between these two channels of energy deposition varies smoothly with CR energy near the stopping power peak. In contrast, for the vast majority of experiments on UV irradiation of ice analogs, the utilized sources emit photons with energies below 11~eV \citep[see][and references therein]{Oeberg2016}. This value is smaller than the ionization threshold of most simple molecules constituting the initial ice matrix \citep{vanDishoeck2006}, and therefore the ion production is usually negligible. Such a choice of UV sources reflects the fact that the interstellar UV field in dense cores is drastically attenuated, and UV photons are generated locally via CR-induced H$_2$ fluorescence \citep{Prasad1983, Cecchi-Pestellini1992}, whose spectrum in molecular gas is dominated by energies around 7.5--8~eV. On the other hand, x-ray photons, only present in molecular material exposed to radiation from young stellar objects \citep[as in case of the inner portions of protoplanetary disks, e.g.,][]{Glassgold1997}, mostly lead to ionization -- ejecting photo- and Auger electrons, which then lose energy to produce further ionization as well as excitation \citep{Oeberg2016, MunozCaro2019}.

Comparison of photolysis products forming in ice upon UV and x-ray irradiation reveals profound differences \citep[see][and references therein]{MunozCaro2019}, which can be attributed to the above differences in the dominant interaction mechanisms. Remarkably, preliminary analysis of the radiolysis products detected in our experiments suggests new striking features in the product abundances and their evolution with fluence, that have not been observed in photolysis experiments. We will report on detailed analysis of the radiolysis chemistry in a separate paper.

\section{Conclusions}
\label{conclusions}

In this paper we show that the scaling dependencies on the stopping power, often used in the literature to describe the sputtering yield and destruction cross section of ice molecules, clearly break down near the peak of $S_{\rm el}(E_{\rm ion})$. The measured destruction cross section of CO deviates from the predicted $\sigma_{\rm des} \propto S_{\rm el}^{1.5}$ dependence, showing a significant enhancement at $E_{\rm ion}$ below the peak. A similar, but less significant deviation from the $Y_{\rm sp}\propto S_{\rm el}^2$ dependence is observed for the sputtering yield. These results suggest that among the two contributions to the electronic stopping power -- ionization and electronic excitations -- the latter is a notably more efficient channel for CO radiolysis. The same conclusion may certainly be correct for CO sputtering, too, However, an interplay of different microscopic processes, possibly involved in the observed sputtering enhancement, cannot be excluded. Further experiments and theoretical work are needed to fully understand the sputtering results.

We speculate that excitation of dissociative electronic states, simultaneously leading to destruction of molecules and acceleration of dissociation products, may be the microscopic process behind the observed results. We also discuss other potential microscopic processes that have been previously mentioned in the literature, and rule out some of them for our experiments. In particular, production of ionized molecules along the track of a projectile ion is shown not to have a noticeable effect. Furthermore, the charge of incident ions has no impact on the destruction cross section, and may have only a marginal effect, less than 10\%, on the sputtering yield. In analysing this latter result we conclude that the actual sputtering depth for CO ice may significantly exceed previous estimates. Additional experiments involving different methods to probe the sputtering depth are needed to verify or rule out this preliminary conclusion.

To the best of our knowledge, the reported experiments are the first dedicated attempt to explore the relative importance of the excitation and ionization channels in the CR processing of astrophysical ices. Our results provide constraints for available physical models of ices undergoing CR bombardment, by examining the efficiency of energy deposition in ice, and also shed light into the leading mechanisms of radiolysis and, thus, improve overall predictability of existing astrochemical models. In a separate publication, we will report on detailed analysis of the radiolysis chemistry driven in the experiments discussed in the present paper.

\vspace{\baselineskip}
We thank Giuseppe Baratta, Guillermo Muñoz Caro, Maria Elisabetta Palumbo, and Kedron Silsbee for useful discussions, Anna Ivleva for assistance with preparing Figure~1, and anonymous referees for their constructive suggestions. The authors gratefully acknowledge funding from the Europlanet 2024 RI which has been funded by the European Union Horizon 2020 Research Innovation Programme under grant agreement No.~871149. The main components of the ICA setup were purchased using funding obtained from the Royal Society through grants UF130409, RGF/EA/180306, and URF/R/191018. Recent developments of the installation were supported in part by the Eötvös Loránd Research Network through grants ELKH IF-2/2019 and ELKH IF-5/2020. This work has also been supported by the European Union and the State of Hungary; co-financed by the European Regional Development Fund through grant No. GINOP-2.3.3-15-2016-00005. We also acknowledge support from the National Research, Development, and Innovation Fund of Hungary through grant No.~K128621. Alexei V. Ivlev, Barbara M. Giuliano, and Paola Caselli acknowledge the Max Planck Society for financial support. Zoltán Juhász is grateful for the support of the Hungarian Academy of Sciences through the János Bolyai Research Scholarship. Duncan V. Mifsud is the grateful recipient of a University of Kent Vice-Chancellor’s Research Scholarship. Sergio Ioppolo acknowledges the Royal Society for financial support. We would like to extend our thanks to the technical and operational staff at the Tandetron Laboratory at Atomki for their continued assistance.

\appendix

\section{Appendix A: Local and average stopping power}
\label{A1}

To simplify the analysis, we consider a situation where the distribution of charge states of the projectile ions in ice is dominated by one or by two neighboring states. For the equilibrium charge distribution of He ions, this is the case for energies (i) $E_{\rm ion}\gtrsim0.6$~MeV, where the neutral-state fraction falls below 5\%, and (ii) $E_{\rm ion}\lesssim0.2$~MeV, where doubly charged ions are below 5\% \citep{Pivovar1962, Bug2013}.

Let us consider regime (i). Fractions $f_1$ and $f_2$ of the singly- and doubly-charged states for a given $E_{\rm ion}$ are functions of the column density $N$ traversed by ions in ice. They are described by the following rate equation:
\begin{eqnarray*}
  \frac{df_1}{dN} = -\sigma_{12}f_1+\sigma_{21}f_2,\label{balance1} \\
  f_1+f_2= 1,\label{balance2}
\end{eqnarray*}
where $\sigma_{ij}$ is the cross section of charge changing from state $i$ to state $j$. For singly charged incident ions, $f_1(0)=1$, we have
\begin{equation}\label{f1}
f_1(N)=\frac{\sigma_{21}}{\sigma_\Sigma}+\frac{\sigma_{12}}{\sigma_\Sigma}e^{-\sigma_\Sigma N},
\end{equation}
and $f_2=1-f_1$, where $\sigma_\Sigma\equiv \sigma_{12}+\sigma_{21}$.

The local stopping power is
\begin{equation}\label{S_loc}
S(N)=S_1f_1(N)+S_2f_2(N),
\end{equation}
where $S_1$ and $S_2>S_1$ are the stopping power values for singly- and doubly-charged ions. This allows us to calculate the local deviation $S(N)-S_{\rm el} \equiv \Delta S(N)$ from the equilibrium stopping power
\begin{equation}\label{S_el}
S_{\rm el}=S_1\frac{\sigma_{21}}{\sigma_\Sigma}+S_2\frac{\sigma_{12}}{\sigma_\Sigma}\;.
\end{equation}
We obtain for the relative deviation:
\begin{equation}\label{N_S}
\frac{\Delta S(N)}{S_{\rm el}}= -\left(\frac{S_2/S_1-1}{S_2/S_1+\sigma_{21}/\sigma_{12}}\right)e^{-\sigma_\Sigma N},
\end{equation}
where $\sigma_{21}/\sigma_{12}=(f_1/f_2)_{\rm eq}\lesssim1$ in regime (i). As $S_2/S_1$ noticeably exceeds unity, we have $\Delta S(N)/S_{\rm el}\sim -e^{-\sigma_\Sigma N}$.

The average stopping power along the track $N$ is given by $\bar S(N)=N^{-1}\int_0^NS(N')dN'$. Substituting Equations~(\ref{f1}) and (\ref{S_loc}), we obtain
\begin{equation}\label{S_eff}
\bar S(N)=S_{\rm el}-\frac{(S_2-S_1)\sigma_{12}}{\sigma_\Sigma^2N} \left(1-e^{-\sigma_\Sigma N}\right),
\end{equation}
and the relative deviation of the average stopping power from equilibrium is given by
\begin{equation}\label{N_Sbar}
\frac{\Delta \bar S(N)}{S_{\rm el}}= -\left(\frac{S_2/S_1-1}{S_2/S_1+\sigma_{21}/\sigma_{12}}\right)
\frac{1-e^{-\sigma_\Sigma N}}{\sigma_\Sigma N}\;.
\end{equation}
Unlike Equation~(\ref{N_S}), this decreases approximately as $\Delta \bar S(N)/S_{\rm el}\sim -(\sigma_\Sigma N)^{-1}$.

The above results can be straightforwardly applied to regime (ii) dominated by neutral atoms and singly charged ions, with the respective fractions $f_0$ and $f_1$, and the stopping powers $S_0$ and $S_1>S_0$. For the incident beam of singly charged ions, it is done by the subscript replacement $2\to0$ in Equations~(\ref{f1})--(\ref{N_Sbar}).

\bibliographystyle{apj}
\bibliography{refs}

\end{document}